\documentclass[a4paper,prd,preprint,tightenlines,floatfix,preprintnumbers,nofootinbib,eqsecnum]{revtex4}

\pdfoutput=1

\usepackage{mathtools}
\usepackage{slashed,color}
\usepackage{hyperref}

\usepackage[absolute,overlay]{textpos} 

\begin{document}

\begin{textblock*}{3cm}(17.5cm,.5cm) 
    \noindent\includegraphics[width=3cm]{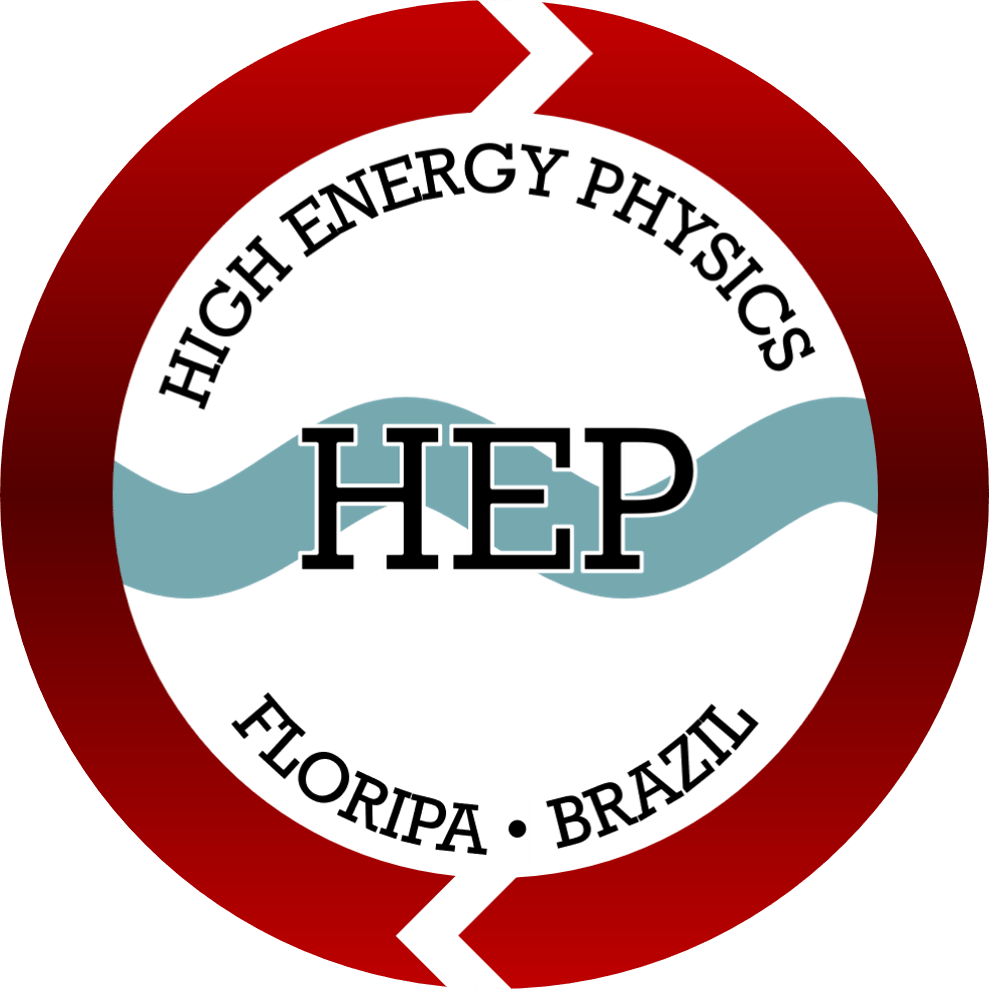} 
\end{textblock*}

\title{Dissociative associated $J/\psi$ and dimuon production \\ in $Ap$ ultraperipheral collisions via double parton scattering}

 \author{Bruna O. Stahlh\"ofer}
\email{stahlhoferbruna@gmail.com}

\author{Edgar Huayra}
\email{yuberth022@gmail.com}
 
\author{Emmanuel G. de Oliveira}
\email{emmanuel.de.oliveira@ufsc.br}

\affiliation{
Departamento de F\'isica, CFM, Universidade Federal de Santa Catarina,
Florian\'opolis, Santa Catarina, Brazil
}

\begin{abstract}
We study the dissociative associated production of a $J/\psi$ meson and a dimuon via double parton scattering (DPS) in nucleus--proton ultraperipheral collisions. This new channel, characterized by a rapidity gap, is sensitive to the photon and gluon distributions of the proton. We derive a DPS pocket formula for this process, together with a corresponding expression for the effective cross section. Furthermore, we demonstrate the kinematic dependence of the effective cross section and present predictions for the differential DPS cross section at LHC and FCC energies.
\end{abstract}


\vspace*{.5cm}
\maketitle

\section{Introduction}
\label{Sec:intro}

In high-energy hadron collisions, the hadron structure manifests itself through the interaction of the constituent partons. Multiple parton interactions (MPI) are known to be very common, leading to the production of complex final states and influencing the distribution of secondary particles~\cite{Paver:1982yp, Mekhfi:1983az, Sjostrand:1987su, Diehl:2011yj}. A particular example of MPI is the double parton scattering (DPS), which occurs between two partons (typically quarks or gluons) of a projectile hadron and two partons of the target hadron. The subject has been extensively measured and explored in the literature, for instance in Refs.~\cite{Bartalini:2011jp, Blok:2010ge, Manohar:2012pe, Diehl:2017kgu, Bansal:2014paa, Huayra:2023gio, Lovato:2025jgh}.

The DPS is sensitive to correlations between pairs of partons, especially in the impact parameter distribution. Within the naive factorized model, in which the DPS cross section is the product of two single parton scattering (SPS) cross sections, an effective cross section appears in the denominator, encapsulating any residual correlations. This parameter has been mostly measured in $pp$ collisions. The case of $pA$ collisions is also drawing attention~\cite{Blok:2012jr,dEnterria:2017yhd, Bashir:2022ufe}, with some effective cross section measurements~\cite{LHCb:2020jse, CMS:2024wgu}. Double parton interactions initiated by photons in $\gamma p$ and $\gamma A$ collisions have also been investigated~\cite{Ceccopieri:2021luf, Blok:2025kvs}, and the respective effective cross sections have been calculated. 

Our interest here is to advance the study of DPS in $Ap$ ultraperipheral collisions (UPCs). This kind of collision is different from the standard $pp$ DPS and features the photon distribution of hadrons. This was first explored in Ref.~\cite{Huayra:2019iun}, with the calculation of the effective cross section involving two real photons from the nucleus interacting with two gluons from the proton to produce $c\bar{c} b \bar{b}$ in $Ap$ UPCs, without projectile nucleus break-up. The calculation was later extended to the case of $AA$ UPCs in Ref.~\cite{Huayra:2020iib}. In a third work \cite{Huayra:2021eve}, we investigated the associated production of $c \bar{c}$ and $l^+ l^-$ in $AA$ UPCs, taking a gluon and a photon from the target nucleus. In these three cases, the effective cross section is not a constant but depends on the longitudinal momenta of the photons. Recently, there has been an interesting measurement of coincident dimuon and $\rho^0$ photon-initiated production in $AA$ coherent UPCs~\cite{ATLAS:2025nac}.

\begin{figure}[t]
    \centering
    \includegraphics[width=.8\textwidth]{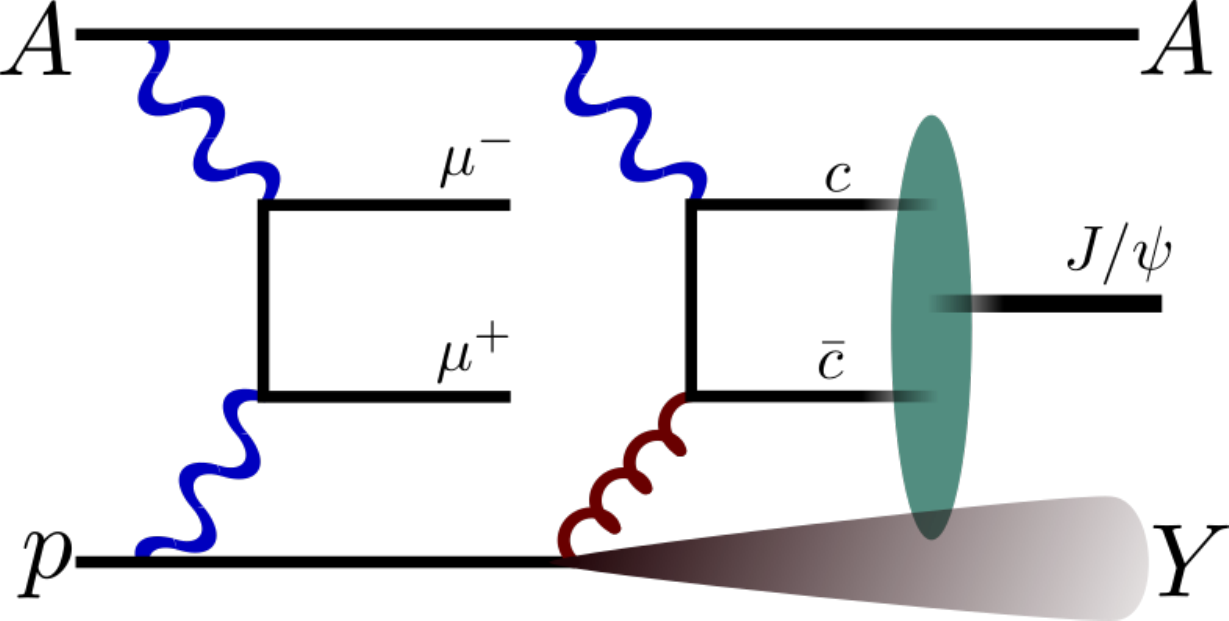}
    \caption{Illustration of the proposed process: DPS in $Ap$ UPC with proton dissociation. The initial state consists of three photons and one gluon, producing a $\mu^-\mu^+$ pair and a charm quark pair $c\bar{c}$ that subsequently hadronizes into a $J/\psi$ meson. The projectile nucleus $A$ remains intact after the collision.}
    \label{fig:esquematico_DPS}
\end{figure}

Continuing our work, in this paper we study a proton dissociative process, characterized by a rapidity gap, with the nucleus acting as a source of photons in an ultraperipheral collision. We introduce here the idea of the associated production of a $J/\psi$ meson and a dimuon. Both exclusive dimuon and dissociative $J/\psi$ production were recently measured by ALICE~\cite{ALICE:2023mfc}. We therefore wonder if both can be measured together in a DPS event. In this case, the parton initial state is composed of $\gamma \gamma \gamma g$, as shown in Fig.~\ref{fig:esquematico_DPS}. This coincident production is rarer, and its measurement may only be feasible with the increased statistics available in Run~3 or, ultimately, at the HL-LHC.

Since we have an ultraperipheral collision, the photon flux from the nucleus — our projectile that remains intact after the collision — is obtained through the equivalent photon approximation (EPA). For the dissociative $J/\psi$ production, which involves a gluon from the target proton, we employ the Color Evaporation Model (CEM) and fit it to H1, ZEUS, and ALICE data. We also need an elastic photon from the target for the dimuon production, considering the contribution of photons inside the proton as well; i.e., we require the impact-parameter-dependent photon distribution. 

With the above ingredients, we derive an analogue of the standard DPS pocket formula for the observable considered in this paper. We calculate the corresponding effective cross section, which strongly depends on the energy fraction of each photon in the initial state, unlike in central collision scenarios. We then evaluate the DPS differential cross section as a function of $J/\psi$ and dimuon rapidity. By comparing the rapidity dependence of the SPS and DPS differential cross sections, we observe that the DPS result differs from the SPS one; i.e., the DPS case is not merely the SPS result times a constant factor. This is a consequence of the nonconstant effective cross section and will help to identify DPS production in future measurements.

This article is organized as follows. In Section~\ref{Sec:inside}, we discuss the parton distributions, i.e., the elastic photon and the dissociative gluon distributions of the target proton, as well as the elastic photon flux of the projectile nucleus. In Section~\ref{Sec:dimuon}, we review exclusive dimuon production in $Ap$ UPCs. In Section~\ref{Sec:Jpsi}, we provide an overview of $J/\psi$ production in $\gamma p$ collisions and present a CEM fit for the dissociative case. In Section~\ref{Sec:predictions}, we develop the main formula for the dissociative associated $J/\psi$ and $\mu^-\mu^+$ production in $Ap$ UPCs; i.e., after the collision, there is a rapidity gap, the projectile nucleus survives intact, and the target proton dissociates. We also numerically calculate the effective cross section and present predictions for this associated production at $\sqrt{s} = 8.16$\,TeV and at $\sqrt{s} = 62.8$\,TeV. We present our conclusions in Sec.~\ref{Sec:concl}.

\section{Photon flux and parton distribution function}
\label{Sec:inside}

In ultraperipheral collisions (UPCs), hadrons interact electromagnetically, since photons are partons that can be found farther away from the hadron than gluons and quarks. It is customary to establish that the collision impact parameter $b$ must exceed the sum of the radii of the colliding hadrons; in the $Ap$ case, $b > R_A + R_p$. Our interest is DPS; therefore, the nucleus, having a larger electric charge, contributes with two elastic photons, as this process is enhanced by the fourth power of the atomic number, $Z^4$. On the proton side, one photon and one gluon are considered, leading to proton dissociation. In the following subsections, we detail the parton distributions (or photon fluxes) employed in our calculations.
\subsection{Proton photon flux}

The photon flux is described using the equivalent photon approximation (EPA), also known as the Weizs\"acker–Williams method~\cite{Fermi:1924tc, vonWeizsacker:1934nji, Williams:1934ad, Williams:1935dka,Jackson:1998nia}, developed in detail in Ref.~\cite{Budnev:1975poe} and summarized, for instance, in Ref.~\cite{Harland-Lang:2018iur}. The flux generated by the proton is expressed as a function of the photon longitudinal momentum fraction $\xi$:
\begin{align}
	\overline{N}^p (\xi) = \frac{d N^p}{d \xi} = 
	\frac{\alpha}{\pi^2 \xi}
	\int \frac{d^2 q_t}{q_t^2+\xi^2 m_p^2}
        \bigg(
	\frac{q_t^2 (1-\xi)}{q_t^2+\xi^2 m_p^2} F_E(Q^2) +
	\frac{\xi^2}{2}F_M(Q^2)
        \bigg).
\label{eq:Nxi}
\end{align}
Here, $\alpha$ is the fine-structure constant, $m_p = 0.938$~GeV is the proton mass, $q_{t}$ denotes the photon transverse momentum, and the photon virtuality is given by
\begin{align}
    Q^2 = \frac{q_t^2 + \xi^2 m_p^2}{1-\xi}.
\end{align}

The functions $F_E$ and $F_M$ are the elastic form factors denoted as $D(Q^2)$ and $C(Q^2)$, respectively, in Tab.~8 of Ref.~\cite{Budnev:1975poe}. They can be written in terms of the Sachs form factors $G_E(Q^2)$ and $G_M(Q^2)$~\cite{Rosenbluth:1950yq, Hofstadter:1958psf, Galynsky:2012dp}:
\begin{align}
    F_E(Q^2) = \frac{4m_p^2 G^2_E(Q^2) + Q^2 \mu^2_p G^2_M(Q^2)}{4m_p^2+Q^2}, 
    \qquad 
    F_M(Q^2) = \mu^2_p G^2_M(Q^2),
\end{align}
where $\mu^2_p = 7.78$ is the square of the proton magnetic moment. For the Sachs factors, we employ the standard dipole parameterization
\begin{align}
       G_E(Q^2) = G_M(Q^2) \equiv \frac{1}{\left(1+Q^2/(0.71\,\text{GeV}^2)\right)^{2}}.
\end{align}
The integration over all $q_t$ in Eq.\ \ref{eq:Nxi} is strongly suppressed at large $q_t$ (or large $Q^2$) by the form factors.

We require the photon distribution in all regions, including photons near the proton boundary or even inside the proton. To achieve this, it is necessary to determine the photon flux as a function of the transverse distance from the hadron center, $\vec{b}_\gamma$. The equivalent photon approximation was extended to this case in Ref.~\cite{Krauss:1997vr} for nuclei and more recently studied for the proton in Ref.~\cite{Dyndal:2015ozg}. 

Following Ref.~\cite{Harland-Lang:2018iur}, we neglect the $F_M(Q^2)$ contribution, as it is multiplied by $\xi^2$ and is therefore very small, a fact we explicitly tested in our observables. With our notation and virtuality definition, the $\xi$- and $\vec{b}_\gamma$-dependent photon flux is given by:
\begin{align}
N^p(\xi,\vec{b}_\gamma) & = \frac{d N^p}{d\xi d^2\vec{b}_\gamma}
= \frac{\alpha}{\pi^2 \xi} \Bigg| \int_0^\infty d q_t \frac{q_t^2}{q_t^2 + \xi^2 m_p^2} \left[ (1-\xi) F_E(Q^2) \right]^{1/2} 
J_1(b_\gamma q_t) 
\Bigg|^2.
\label{eq:Nxib}
\end{align}
It can be verified that integrating Eq.~\ref{eq:Nxib} over $\vec{b}_\gamma$ reproduces Eq.~\ref{eq:Nxi}, apart from the neglected $F_M(Q^2)$ contribution.

\begin{figure}[tb]
\centering
\includegraphics[width=.49\textwidth]{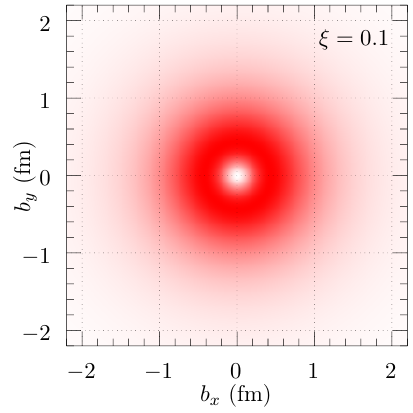}
\includegraphics[width=.49\textwidth]{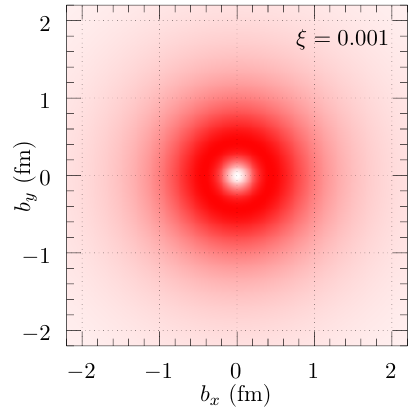}
\caption{Photon distributions of the proton on the impact-parameter $b_{\gamma}$ plane for two values of the momentum fraction $\xi$. On the left, the distribution for $\xi = 0.1$ is more concentrated, while on the right, for $\xi = 0.001$, it is more spread out. In both cases, the distributions peak at $b_{\gamma} \sim 0.5$ fm, showing that the dominant photon contribution originates inside the proton.}
\label{fig:N_barXi_b_real}
\end{figure}

Fig.~\ref{fig:N_barXi_b_real} shows how photons are distributed both inside and outside the proton, given by the single dipole parameterization. The distribution exhibits a peak around $b_\gamma \approx 0.5$\,fm which, compared to the proton radius (0.84\,fm), indicates that most photons are located inside the proton. It also illustrates the dependence on the photon momentum fraction: for $\xi = 0.1$ (left), photons are more concentrated, whereas for $\xi = 0.001$ (right), they are more spread out. 

We also test a double dipole parameterization
\begin{align}
	G_{E,M} (Q^2) = \frac{a_0^{E,M}}{\left(1+Q^2/a_1^{E,M}\right)^2} + \frac{1-a_0^{E,M}}{\left(1+Q^2/a_2^{E,M}\right)^2},
\end{align}
which was fitted to experimental data by the A1 Collaboration~\cite{A1:2013fsc} and describes them surprisingly well. The parameters are listed in Table~\ref{tab:double-dipole_coef} and in Ref.~\cite{Bernauer:2010zga}. The results in Fig.~\ref{fig:N_barXi_b_real} and subsequent figures change very little when replacing the single dipole by the double dipole parameterization. Therefore, although we repeated all calculations with the double dipole parameterization, we do not show them in this paper. 

\begin{table}[t]
\centering
\begin{tabular}{ |c|c|c|c|  }
  \hline
       & $a^0$    & $a^1$    & $a^2$ \\
 \hline
 $G_E(Q^2)$   &  0.98462 &  0.68414 & 0.01933 \\
 $G_M(Q^2)$   &  0.28231 &  1.34919 & 0.55473 \\
 \hline
\end{tabular}
\caption{Parameters of the Sachs form factors, both electric and magnetic, using the double dipole parameterization fitted to data~\cite{A1:2013fsc}.} 
\label{tab:double-dipole_coef}
\end{table}

\subsection{Gluon distribution function of the proton}

In a simple approach, the impact-parameter-dependent gluon distribution $G_g(x,\vec{b}_g)$ is factorized as
\begin{align}
G_g(x,\vec{b}_g) = g(x)\, f_g(\vec{b}_g).
\end{align}
We use a collinear parton distribution function (PDF) $g(x)$ for the gluons, namely the CT18LO set~\cite{Yan:2022pzl}. There is an implicit dependence on factorization scale in this function, that will be discussed in the next section. 

For the transverse profile $f_g(\vec{b}_g)$, we employ the following function (Ref.~\cite{Frankfurt:2010ea}, e.g.):
\begin{align}
f_g (\vec{b}_g) = \frac{\Lambda^2}{2\pi} \frac{\Lambda b_g}{2} K_1 (\Lambda b_g),
\end{align}
which is normalized such that $\int d^2 \vec{b}_g\, f_g (\vec{b}_g) = 1$, with $\Lambda = 1.5$~GeV.

\subsection{Nuclear photon flux}

For the projectile nucleus, we adopt the approximation of a photon flux from a point-like charge, since the interactions occur outside the nucleus. The nucleus remains intact, and therefore these are elastic photons. The distribution in momentum fraction $\xi$ and transverse distance from the nucleus center $b_\gamma$, valid for photons outside the nucleus ($b_\gamma > R_A$), is given by
\begin{align}
     N^A(\xi, \vec{b}_\gamma) = \frac{d N^A}{d\xi d^2 \vec{b}_\gamma} 
     = \frac{Z^2 \alpha k^2}{\pi^2 \xi b_\gamma^2} 
     \left[K_1^2(k) + \frac{1}{\gamma^2}K_0^2(k) \right],
\end{align}
where $k = \xi b_\gamma m_p$. This expression is reviewed in Ref.~\cite{Vogt:2007zz}. The longitudinal momentum fraction is
\begin{align}
\xi = \frac{2\omega}{\sqrt{s}} = \frac{\omega}{\gamma m_p},
\end{align}
where $\omega$ is the photon energy, $\gamma$ the Lorentz factor, and $\sqrt{s}$ the center-of-mass energy per nucleon.

\section{Exclusive dimuon production in UPCs}
\label{Sec:dimuon}

\begin{figure}[tb]
\centering
\includegraphics[width=.70\textwidth]{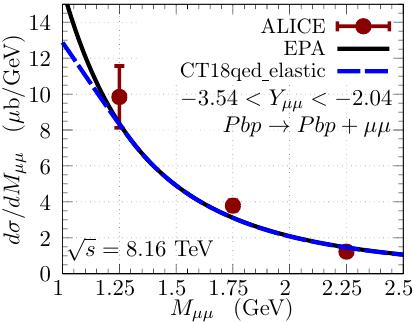}
\caption{Differential cross section for dimuon ($\mu^- \mu^+$) production in $Pbp$ UPCs at the LHC energy of $\sqrt{s} = 8.16$ TeV as a function of the invariant mass, with the dimuon rapidity $Y_{\mu\mu}$ integrated over. Our theoretical calculations, labeled as ``EPA'', are compared with experimental data from the ALICE Collaboration~\cite{ALICE:2023mfc}, showing good agreement. We also include a comparison with results obtained using the ``CT18qed\_elastic'' parton distribution function.}
\label{fig:SPSdilepton}
\end{figure}

The SPS differential cross section for exclusive dimuon production in $Ap$ UPCs 
is given by
\begin{align}
\frac{d^2\sigma_{Ap \rightarrow Ap \mu \mu}}{dY_{\mu \mu} dM^2_{\mu \mu}} 
& = \int d^2 \vec{b} \, \Theta(b - R_A - R_p) \, \int d^2\vec{b}_{\gamma_1} \, \nonumber \\
&\times 
\frac{\xi_1 \xi_2}{M^2_{\mu \mu}} \, \Theta(b_{\gamma_1} - R_{A}) \, N^A(\xi_1,\vec{b}_{\gamma_1}) \, N^p(\xi_2,\vec{b}_{\gamma_1} - \vec{b}) \hat{\sigma}_{\gamma \gamma \rightarrow \mu \mu}.
\label{eq:SPSdimuon}
\end{align}
Here, $Y_{\mu \mu}$ and $M_{\mu \mu}$ denote the dimuon rapidity and invariant mass, respectively, in the nucleon--proton center-of-mass frame. The momentum fractions are given by 
\begin{align}
\xi_{1,2} = \frac{M_{\mu \mu}}{\sqrt{s}} \mathrm{e}^{\pm Y_{\mu \mu}}.
\end{align}
The Heaviside function $\Theta(b-R_A-R_p)$ ensures ultraperipheral collisions. The partonic $\gamma \gamma \rightarrow \mu^-\mu^+$ Breit--Wheeler cross section~\cite{Breit:1934zz, Brodsky:1971ud} reads
\begin{align}
\hat{\sigma}_{\gamma \gamma \rightarrow \mu \mu} = 
\frac{4 \pi \alpha^2}{M_{\mu\mu}^2} 
\left\{\left[2 + 2\beta - \beta^2\right] 
\ln \left(\frac{1}{\beta} + \sqrt{\frac{1}{\beta} - 1}\right) - 
\sqrt{ 1 - \beta} \left(1 + \beta \right)\right\},
\label{eq:SPSdilepton3}
\end{align}
with $\beta = 4m^2_{\mu}/M_{\mu\mu}^2$ and $m_\mu$ the muon mass.
 
In Fig.~\ref{fig:SPSdilepton}, we present the ALICE measurement of exclusive dimuon photoproduction in ultraperipheral $Ap$ collisions at $\sqrt{s} = 8.16$\,TeV, within the rapidity range $-3.54 < Y_{\mu\mu} < -2.04$. We work in the nucleon--proton center-of-mass frame and adopt the convention in which the nucleus moves from left to right and the proton from right to left. In contrast, ALICE reports data in the laboratory frame with the opposite convention. We have performed the necessary adjustments to their data for comparison in Fig.~\ref{fig:SPSdilepton}. 

Our theoretical result using photons from Eq.~\ref{eq:Nxi} with the single-dipole parameterization is shown as the ``EPA'' black solid curve in Fig.~\ref{fig:SPSdilepton}, exhibiting good agreement with the data. We also show the ``CT18qed\_elastic'' result (blue dashed curve), obtained using photons from the CT18qed PDFs~\cite{Xie:2023qbn}, which confirms that there is no significant discrepancy between the two calculations.

\section{Dissociative $J/\psi$ production in UPCs}
\label{Sec:Jpsi}

To evaluate the dissociative SPS $J/\psi$ production, we use the Color Evaporation Model (CEM)~\cite{Fritzsch:1977ay,Halzen:1977rs}, which provides a framework for obtaining $J/\psi$ hadronization from charm--anticharm pair production. The $c\bar{c}$ cross section in $\gamma p$ collisions is integrated over the $c\bar{c}$ energy range from the $J/\psi$ mass, $m_{J/\psi}= 3.10$\,GeV, up to twice the $D$-meson mass, $2m_D = 3.728$\,GeV, and multiplied by the hadronization factor $F_\psi$:
\begin{align}
	\sigma_{\gamma p \rightarrow J/\psi + X} =
	F_{\psi}
	\int_{m^2_{J/\psi}}^{4 m^2_D}
	\, \frac{dM^2_{c\bar{c}} }{M^2_{c\bar{c}}}\, 
	x g(x, \mu_F)
   	\hat{\sigma}_{\gamma g \rightarrow c \bar{c}}.
\end{align}
Here, $M_{c\bar{c}}$ is the invariant mass of the $c\bar{c}$ system, $\mu_F$ the factorization scale, and $x=M^2/W_{\gamma p}^2$ the gluon momentum fraction of the proton, with $W_{\gamma p}$ being the center-of-mass energy of the $\gamma p$ interaction.

The partonic cross section for $c\bar{c}$ production is
\begin{align}
	\hat{\sigma}_{\gamma g \rightarrow c\bar{c}} = 
	\frac{2 \pi \alpha \alpha_{s}(\mu_R) e^2_c}{M_{c\bar{c}}^2}
	\bigg[(1+\beta-\tfrac{1}{2}\beta^2)\log \bigg(\frac{1+\nu}{1-\nu} \bigg)
	- (1+\beta)\nu \bigg],
\end{align}
with $\nu = \sqrt{1-\beta}$ and $\beta = (2 m_c / M_{c\bar{c}})^2$. We set the charm-quark mass to $m_c = 1.4$~GeV and its charge to $e_c = 2/3$. For the strong coupling constant $\alpha_s$, the renormalization scale $\mu_R$ is taken equal to the factorization scale $\mu_F$.

\begin{figure}[tb]
\centering
\includegraphics[width=.70\textwidth]{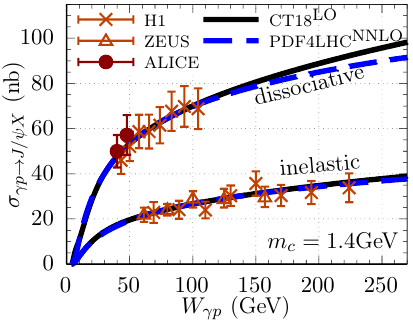}
\caption{Dissociative and inelastic $J/\psi$ production cross sections in $\gamma p$ collisions as a function of the $\gamma p$ c.m. energy $W$. The results are obtained by fitting the CEM to data from ALICE~\cite{ALICE:2023mfc}, ZEUS~\cite{ZEUS:2012qog}, and H1~\cite{H1:2013okq, H1:2010udv}. The CT18LO~\cite{Yan:2022pzl} and NNLO PDF4LHC21~\cite{PDF4LHCWorkingGroup:2022cjn} gluon PDFs were used.}
\label{fig:SimpleJPsi_W}
\end{figure}
\begin{figure}[tb]
\centering
\includegraphics[width=.70\textwidth]{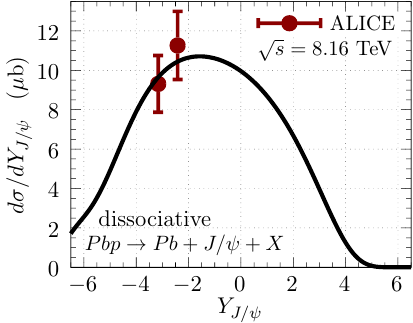}
\caption{SPS dissociative $J/\psi$ production cross section in $Pbp$ UPCs at the LHC ($\sqrt{s} = 8.16$ TeV) as a function of vector meson rapidity $Y_{J/\psi}$, compared with ALICE data~\cite{ALICE:2023mfc}.}
\label{fig:SimpleJPsi_Y}
\end{figure}

We test the model by fitting inelastic $J/\psi$ production to the most recent HERA data~\cite{H1:2010udv, ZEUS:2012qog}. Using the ``Minuit'' minimization algorithm within ``Root''~\cite{James:2004xla}, we obtain $F_{\psi}=0.044 \pm 0.010$ and $\mu_F=2.034 \pm 0.459$\,GeV with the CT18LO gluon distribution. Repeating the procedure with the PDF4LHC21~\cite{PDF4LHCWorkingGroup:2022cjn} NNLO gluon distribution, we find $F_{\psi}=0.058 \pm 0.012$ and $\mu_F=3.911 \pm 1.478$\,GeV. The resulting fits are shown in Fig.~\ref{fig:SimpleJPsi_W} and display good agreement with the data.

Actually, we need the dissociative production. We follow the same procedure, using experimental results from ALICE~\cite{ALICE:2023mfc} and H1~\cite{H1:2013okq}. The parameters obtained are $F_{\psi}=0.125 \pm 0.017$ and $\mu_F=1.910 \pm 0.310$\,GeV with the CT18LO gluon PDF, and $F_{\psi}=0.169 \pm 0.021$ and $\mu_F=3.30 \pm 0.874$\,GeV with the PDF4LHC21 NNLO gluon PDF. These results, shown in Fig.~\ref{fig:SimpleJPsi_W}, demonstrate very good agreement between calculation and data. 

The CEM assumes that one gluon from the proton produces the vector meson, with the extra color ``evaporating''. This is a reasonable assumption for the inelastic case. In the dissociative case, however, a rapidity gap suggests that at least two gluons are exchanged with the proton, forming a color-singlet channel (hard Pomeron). The likely reason for the success of our description is that the color-singlet channel is dominated by a single gluon carrying most of the required momentum fraction, which can be taken from the PDFs. The additional gluons exchanged are essentially soft, contributing little momentum fraction. Their role, beyond enabling the color-singlet final state and the rapidity gap, is effectively absorbed into the fit parameters $F_{\psi}$ and $\mu_F$.

Finally, the differential cross section for dissociative SPS $J/\psi$ production in $Ap$ UPCs is given by
\begin{align} 
\frac{d\sigma_{Ap \rightarrow AX + c\bar{c}}}{dY_{J/\psi}} &= \int d^2 \vec{b} \, \Theta(b - R_A - R_p) \, \int d^2 \vec{b}_{\gamma} \nonumber \\
&\times
F_{\psi} \int^{4m_{D}^2}_{m^2_{J/\psi}} dM_{c \bar{c}}^2 \, \frac{\xi x}{M_{c\bar{c}}^2} \, \Theta(b_{\gamma} - R_{A}) \, N^A(\xi,\vec{b}_{\gamma}) \, G_g(x, \vec{b}_{\gamma} - \vec{b}) \, \hat{\sigma}_{\gamma g \rightarrow c\bar{c}},
\end{align}
where $Y_{c\bar{c}}$ and $M_{c \bar{c}}$ denote the rapidity and invariant mass of the $c\bar{c}$ pair, respectively. The momentum fractions are
\begin{align}
\xi = \frac{M_{c\bar{c}}}{\sqrt{s}} \mathrm{e}^{Y_{c\bar{c}}}, \qquad  
x = \frac{M_{c\bar{c}}}{\sqrt{s}} \mathrm{e}^{-Y_{c\bar{c}}}. 
\end{align} 
In Fig.~\ref{fig:SimpleJPsi_Y}, we show this cross section at the LHC ($\sqrt{s} = 8.16$ TeV) as a function of vector meson rapidity $Y_{J/\psi}$. The comparison with ALICE data~\cite{ALICE:2023mfc} shows good agreement.

\section{DPS cross section results}
\label{Sec:predictions}
	
We want to study the associated $J/\psi\,\mu^-\mu^+$ DPS production in a process where the nucleus (projectile) emits only photons and remains intact, while the proton (target) dissociates after providing both a gluon and a photon. We have described in the previous sections the elements that work reasonably well for SPS production. Therefore, we can use them to pragmatically build the DPS cross section in terms of the parton-level elementary cross sections as
\begin{align}
\frac{d\sigma^{\rm DPS}_{A p}}{dY_{\mu \mu} \, dM^2_{\mu \mu} \, {dY_{J/\psi}}}
&= 
\int d^2 b \, \Theta(b - R_{A} - R_{p}) \int d^2 \vec{b}_{\gamma_1} \, \int d^2 \vec{b}_{\gamma_3} \, \nonumber \\
&\times
\, \frac{\xi_1 \xi_2}{M_{\mu \mu}^2}
\, \Theta (b_{\gamma_1} - R_{A}) \, N^A (\xi_1, \vec{b}_{\gamma_1}) N^p (\xi_2, \vec{b}_{\gamma_1} - \vec{b})
\, \hat{\sigma}^{\mu \mu}_{\gamma \gamma}  \\
&\times 
\, F_{\psi} \int^{4 m^2_{D}}_{m^2_{J/\psi}} d M_{c \bar{c}}^2 \, \frac{\xi_3 x}{M_{c \bar{c}}^2}
\, \Theta (b_{\gamma_3} - R_{A}) \, N^A (\xi_3, \vec{b}_{\gamma_3})\, g(x)
\, f_g(\vec{b}_{\gamma_3} - \vec{b}) \, \hat{\sigma}^{c\bar{c}}_{\gamma g}, \nonumber
\label{eq:DPS_3f1g}
\end{align}
where $\vec{b}_{\gamma_2} = \vec{b}_{\gamma_1}-\vec{b}$ and $\vec{b}_g = \vec{b}_{\gamma_3}-\vec{b}$. This expression neglects correlations between the two projectile photons as well as correlations between gluons and photons in the target.

To encapsulate the impact-parameter dependence, we define the overlap functions $T_{\gamma \gamma} (\xi_1, \xi_2, b)$ and $T_{\gamma g} (\xi, \vec{b})$:
\begin{align}
T_{\gamma \gamma} (\xi_1, \xi_2, \vec{b}) = \frac{1}{\overline{N}^A(\xi_1)\overline{N}^p(\xi_2)} \, \int d^2 \vec{b}_{\gamma_1} \, \Theta(b_{\gamma_1} - R_{A}) N^A(\xi_1,\vec{b}_{\gamma_1}) \, N^p(\xi_2,\vec{b}_{\gamma_1} - \vec{b})
\end{align}
and
\begin{align}
T_{\gamma g} (\xi_3, \vec{b}) = \frac{1}{\overline{N}^A(\xi_3)} \, \int d^2 \vec{b}_{\gamma_3}  \,
\Theta(b_{\gamma_3} - R_{A}) N^A(\xi_3,\vec{b}_{\gamma_3}) f_g(\vec{b}_{\gamma_3} - \vec{b}) \,.
\end{align}

We also define the $\gamma \gamma \to \mu^- \mu^+$ auxiliary subprocess cross section
\begin{align}
\frac{d^2\Sigma^{\mu \mu}_{\gamma \gamma} (\xi_1, \xi_2)}{dY_{\mu \mu}dM^2_{\mu \mu}} = \frac{\xi_1 \xi_2}{M_{\mu \mu}^2} \, \overline{N}^A(\xi_1) \, \overline{N}^p(\xi_2) \, \hat{\sigma}_{\gamma \gamma \rightarrow \mu \mu},
\label{eq:SPSdilepton2}
\end{align}
and the $\gamma g \to J/\psi$ auxiliary subprocess cross section
\begin{align}
\frac{d\Sigma_{\gamma g}^{J/\psi} (\xi_3, x)}{dY_{J/\psi}} = 
F_{\psi}
\int_{m^2_{J/\psi}}^{4m^2_D} d M^2_{c \bar{c}}
\, \frac{\xi_3 \, x}{M_{c\bar{c}}^2} \, \overline{N}^A(\xi_3) \, g(x) \, \hat{\sigma}_{\gamma g \rightarrow c\bar{c}}.
\end{align}

With these definitions, and in analogy with the standard DPS pocket formula, the differential DPS cross section can be expressed as the product of the auxiliary subprocess cross sections:
\begin{align}
\frac{d\sigma^{\rm DPS}_{A p}}{dY_{\mu \mu} \, dM^2_{\mu \mu} \, {dY_{J/\psi}} \,d M^2_{c \bar{c}}}
= \, \frac{1}{\sigma_\text{eff}(\xi_1, \xi_2, \xi_3)}
\, \frac{d\Sigma^{\mu \mu}_{\gamma \gamma} (\xi_1, \xi_2)}{dY_{\mu \mu}dM^2_{\mu \mu}}
\, \frac{d\Sigma^{J/\psi}_{\gamma g} (\xi_3, x)}{dY_{J/\psi} d M^2_{c \bar{c}} }.
\label{eq:DPS_pocket_formula}
\end{align}
The new factor is the effective cross section
\begin{align}
\sigma^{-1}_\text{eff}(\xi_1, \xi_2, \xi_3) 
\equiv \int d^2 b \, \Theta(b - R_{A} - R_{p})  T_{\gamma \gamma}(\xi_1, \xi_2, \vec{b}) T_{\gamma g}(\xi_3, \vec{b}) \,,
\label{eq:siga_eff_3f1g}
\end{align}
which is specific to $Ap$ UPCs with proton dissociation and to processes that involve two elastic photons from the nucleus and both an elastic photon and a dissociative gluon from the proton. 

\begin{figure}[tb]
	\centering
	\includegraphics[width=.49\textwidth]{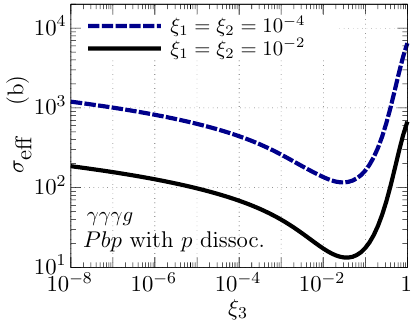}
        \includegraphics[width=.49\textwidth]{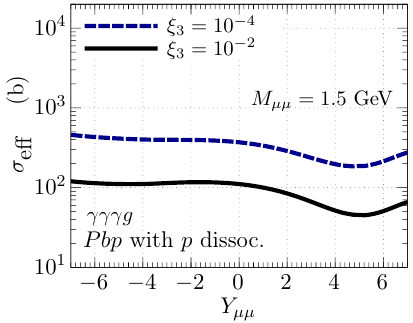}
	\caption{DPS effective cross section in $Ap$ UPCs with proton dissociation considering interactions where two photons from the projectile interact with one gluon and one photon from the target ($\gamma \gamma \gamma g$). Left: dependence on the momentum fraction $\xi_3$ of the photon from the target proton, keeping $\xi_1$ and $\xi_2$ fixed. Right: dependence on $Y_{\mu \mu}$, fixing $\xi_3$ and setting $M_{\mu \mu} = 1.5$ GeV.}
    \label{fig:sigma_eff}
\end{figure}
\begin{figure}[tb]
	\centering
	\includegraphics[width=.7\textwidth]{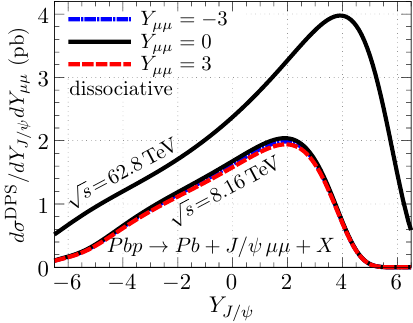}
	\caption{DPS cross section for $J/\psi \, \mu^-\mu^+$ production in dissociative $Pbp$ UPCs at the LHC ($\sqrt{s} = 8.16$ TeV) and FCC ($\sqrt{s} = 62.8$ TeV) as a function of the $J/\psi$ rapidity. The dimuon rapidity $Y_{\mu\mu}$ is fixed at $0$, $3$, and $-3$, with the invariant mass $M_{\mu\mu}$ integrated over. The $Y_{J/\psi}$ dependence is clearly different from the SPS case. The results indicate that the most significant contribution arises for $Y_{\mu\mu} = 0$.}
	\label{fig:DPS_jpsi_muons}
\end{figure}
\begin{figure}[tb]
	\centering
	\includegraphics[width=.7\textwidth]{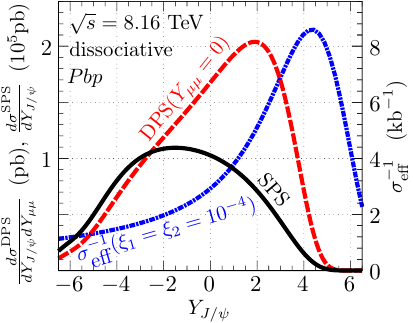}
	\caption{Comparison of DPS ($J/\psi \mu \mu$) and SPS single $J/\psi$ production cross sections, together with the inverse of the effective cross section, in $Pbp$ UPCs at the LHC ($\sqrt{s} = 8.16$ TeV), plotted as functions of $J/\psi$ rapidity. Multiplying the SPS contribution by the inverse of the effective cross section provides an estimate of the expected DPS contribution, apart from a constant related to dimuon production. These results illustrate how the rapidity dependence of the effective cross section influences the behavior of the DPS cross section.}
	\label{fig:DPS_vs_SPS}
\end{figure}

The effective cross section of Eq.~\ref{eq:siga_eff_3f1g} is one of the main results of this paper and is illustrated in Fig.~\ref{fig:sigma_eff}. Its behavior is dominated by the photon longitudinal momenta; of particular importance is the likelihood that the projectile photons $\gamma_1$ and $\gamma_3$ are found together inside or near the target proton, where they can more often interact with the photon $\gamma_2$ and the gluon. In our UPC case, this must occur outside the projectile nucleus.

On the left panel of Fig.~\ref{fig:sigma_eff}, we show the effective cross section with fixed values $\xi_1=\xi_2$ as a function of $\xi_3$, i.e., the momentum fraction of the photon that produces the $J/\psi$. For small $\xi_3$, the effective cross section is larger---and consequently the DPS cross section is smaller---because the $\gamma_3$ photon is too spread out. As $\xi_3$ increases, the effective cross section decreases since it becomes more likely that both $\gamma_3$ and $\gamma_1$ photons interact with the proton partons. The minimum is reached around $\xi_3 \approx 0.03$, where the $\gamma_3$ photon approximately occupies a shell from $r=R_A$ to $R_A+2R_p$. For large $\xi_3$, this shell becomes too thin and the effective cross section increases rapidly, as the photon will often not be found inside the proton (the latter must keep some distance from the nucleus in a UPC).

On the right panel of Fig.~\ref{fig:sigma_eff}, we show the effective cross section with fixed $\xi_3$, fixed $M_{\mu \mu}=1.5$~GeV, as a function of $Y_{\mu \mu}$. In this case, larger effective cross sections are found at negative rapidities, whereas positive rapidities correspond to smaller values. The behavior is dominated by the $\gamma_1$ photon, as the $\gamma_2$ photon from the proton has a distribution that varies less. The minimum is reached at $Y_{\mu\mu} \approx 5$, which corresponds to $\xi_1 \approx 0.03$, as in the previous case.

The presented results were obtained using the single-dipole parameterization of the Sachs form factors. We have tested the double-dipole parameterization as well, and the effective cross section is almost the same. This can be explained by the fact that the nucleus photons that interact with the proton photon to produce dimuons are spread over a large region in impact parameter and therefore are not sensitive to small details of the target. We also verified that both parameterizations, for small $\xi$, yield similar photon distributions. One concludes that there is no significant dependence on the shape of the form factors, which is an advantage for studying gluon–photon correlations.

In Fig.~\ref{fig:DPS_jpsi_muons}, we present the dissociative $Pbp$ DPS $J/\psi\,\mu\mu$ production cross section, differential in $J/\psi$ rapidity, at fixed dimuon rapidities $Y_{\mu\mu}=-3,\,0,\,\text{and }3$ and at the LHC energy $\sqrt{s}=8.16$~TeV. These results are a first prediction for this (as yet unmeasured) observable. We find a weak dependence on the dimuon rapidity, which is itself an experimental advantage. We also observe a very different behavior compared to the SPS case of Fig.~\ref{fig:SimpleJPsi_Y}: the DPS production is dominated by forward $Y_{J/\psi}$, in contrast to the SPS case. In order to gauge the feasibility of a measurement, we estimate fewer than one event for the ALICE integrated luminosity of $7.9\,\mathrm{nb}^{-1}$ collected in 2016 and a handful of events for the Run~3 target ALICE luminosity of $0.25\,\mathrm{pb}^{-1}$.

The key difference between the DPS and SPS results is the effective cross section, which depends on the $J/\psi$ rapidity. This is illustrated in Fig.~\ref{fig:DPS_vs_SPS}, where the SPS and DPS cross sections are plotted together with $\sigma_\text{eff}^{-1}$. By multiplying the SPS contribution by the inverse of the effective cross section, one can estimate the expected DPS contribution, up to a constant factor related to dimuon production. The inverse effective cross section is larger at forward rapidities, and this is why the SPS peak at negative rapidities is shifted to positive rapidities in the DPS production.
\section{Conclusion}
\label{Sec:concl}	

We have investigated double parton scattering in ultraperipheral $Ap$ collisions with proton dissociation, simultaneously probing the photon and gluon content of the proton. In this scenario, the nucleus contributes with two elastic photons, and a particularly sensitive observable is the dissociative production of a $J/\psi$ meson in association with a dimuon. This final state produces a rapidity gap in the detector, providing a clear experimental signature.    

For the SPS dissociative $J/\psi$ production, we employed the Color Evaporation Model (CEM) to evaluate the cross section, using the CT18LO collinear parton distribution. The factorization scale and the hadronization factor were fitted to H1 and ALICE data, yielding good agreement. For exclusive dimuon SPS production, we used the equivalent photon approximation with the dipole parameterization of the proton form factor, also finding consistency with ALICE data.

The associated DPS production of $J/\psi$ and dimuons offers a novel way to probe the photon density inside the proton and its correlation with the gluon density. For simplicity, we assumed no explicit photon–gluon correlations, an approximation partly supported by the distinct distributions expected from QED and QCD dynamics. We derived an analogue of the DPS pocket formula, computed for the first time the corresponding effective cross section, and studied its dependence on the longitudinal momentum fractions of the photons. 

We presented predictions for the differential cross section as a function of $J/\psi$ and dimuon rapidity. At the LHC energy of $\sqrt{s} = 8.16$~TeV, our results indicate a sizable DPS contribution, potentially observable in forthcoming measurements. We also provide predictions at energy of  $\sqrt{s} = 62.8$~TeV. Importantly, the DPS cross section is not simply the product of two SPS cross sections: the momentum-fraction dependence of the effective cross section modifies the rapidity dependence of the $J/\psi$ and dimuon distributions relative to the SPS case. Altogether, our study motivates future experimental investigations of $J/\psi \mu^-\mu^+$ final states in dissociative $Ap$ UPCs.

\section*{Acknowledgments}

This work was supported by FAPESC, INCT-FNA (464898/2014-5), and the National Council for Scientific and Technological Development – CNPq (Brazil) for BOS, EGdO, and EH. This study was also financed in part by the Coordenação de Aperfeiçoamento de Pessoal de Nível Superior -- Brasil (CAPES) -- Finance Code 001.


\appendix

\section{Dissociative double $J/\psi$ DPS production in UPC $Pbp$ collisions}
\label{Sec:JJ}	

\begin{figure}[htb]
	\centering
	\includegraphics[width=.45\textwidth]{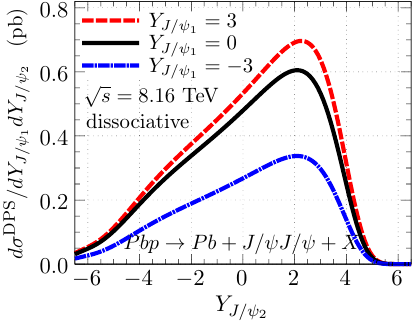}
        \includegraphics[width=.45\textwidth]{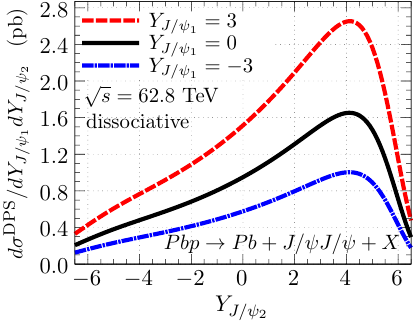}
	\caption{DPS cross section for double $J/\psi$ production in $Pbp$ UPCs as a function of $J/\psi$ rapidity. Left: LHC energy $\sqrt{s}=8.16$~TeV. Right: FCC energy $\sqrt{s}=62.8$~TeV.}
	\label{fig:DPS_Jpsi_Jpsi}
\end{figure}

It is also of interest to calculate the dissociative double $J/\psi$ DPS production in UPC $Pbp$ collisions. For this purpose, we use the effective cross section previously calculated for $\gamma\gamma gg$ initial partons in Ref.~\cite{Huayra:2019iun}. Following the idea of the present analysis, we assume that the transverse distributions of partons are the same in the dissociative and in the inelastic cases. In Fig.~\ref{fig:DPS_Jpsi_Jpsi}, we present the DPS cross section for $J/\psi \, J/\psi$ production, differential in the rapidity of one of the mesons ($Y_{J/\psi_1}$), with the rapidity of the second meson fixed at $Y_{J/\psi_2}=0,\,3,\,\text{and}\,-3$. We observe that the overall behavior is similar to that of the mixed $J/\psi \, \mu^-\mu^+$ channel. However, in this case there is a more pronounced dependence on $Y_{J/\psi_2}$, in contrast to the previous channel. The magnitude of the $J/\psi J/\psi$ DPS cross section is smaller than in the mixed $J/\psi\mu\mu$ production, suggesting that the latter provides a stronger DPS signal.

\end{document}